\documentclass[aps,prb,twocolumn,showpacs, superscriptaddress, floatfix]{revtex4}
\usepackage{graphicx}

\begin{document}
\title{Free-carrier relaxation and lattice heating in photoexcited bismuth thin films}
\author{Y. M. Sheu}
\affiliation{Center for Integrated Nanotechnologies, MS K771, Los Alamos National Laboratory, Los Alamos, New Mexico 87545, USA}
\author{Y. J. Chien}
\affiliation{AU Optronics Corporation, 1 JhongKe Rd. Central Taiwan Science Park,
Taichung 40763, Taiwan}
\author{C. Uher}
\affiliation{Department of Physics, University of Michigan, Ann Arbor, Michigan 48109-1040, USA}
\author{S. Fahy}
\affiliation{Tyndall National Institute and Department of Physics, University College, Cork, Ireland}
\author{D. A. Reis}
\affiliation{PULSE Institute, SLAC National Accelerator Laboratory Menlo Park, CA, 94025 USA}
\affiliation{Departments of Applied Physics and Photon Science, Stanford University, Stanford, CA 94305 USA}


\begin{abstract}
We report ultrafast surface pump and interface probe experiments on photoexcited carrier transport across single crystal bismuth films on sapphire.   The film thickness is sufficient to separate  carrier dynamics from  lattice heating and strain, allowing us to investigate the time-scales of momentum relaxation, heat transfer to the lattice and electron-hole recombination.    The measured electron-hole ($e-h$) recombination time is 12--26 ps and ambipolar diffusivity is 18--40  cm$^{2}$/s for carrier excitation up to $\sim 10^{19}~\text{cm}^{-3}$.  By comparing the heating of the front and back sides of the film, we put lower limits on the rate of heat transfer to the lattice, and by observing the decay of the plasma at the back of the film, we estimate the timescale of electron-hole recombination. We interpret each of these timescales within a common framework of electron-phonon scattering  and find qualitative agreement between the various relaxation times observed. We find that the carrier density is not determined by the $e-h$ plasma temperature after a few picoseconds. The diffusion and recombination become nonlinear with initial excitation $\gtrsim 10^{20}~\text{cm}^{-3}$.
\end{abstract}
 \maketitle

\section{Introduction}

Intense ultrafast laser excitation of solids can alter dramatically the interatomic forces by depositing substantial amounts of energy into free-carriers on a time scale that is short compared to energy relaxation. The resulting non-thermal distribution of electrons in the bands of the material can drive coherent phonon motion, as well as dissipate its energy to the lattice by incoherent electron-phonon scattering. The evolution of the interatomic forces following photoexcitation is strongly dependent on the evolution of the electron distribution as it relaxes back to thermal equilibrium.

Free carrier relaxation in metals, semimetals and semiconductors is strongly influenced by the electronic density of states at the energies where electrons and holes are present.
Near equilibrium, electrons and holes in both metals and semimetals lie in an energy range of a few $kT$ of the Fermi energy ($k$ is the Boltzmann constant and $T$ is the absolute temperature).
The significant difference between the two is in the density of states at the Fermi level: for a typical metal, this is of the order of 1~eV$^{-1}$ (per unit cell); for a semimetal like bismuth, it is of the order of $10^{-3}$~eV$^{-1}$.
In metals, the relatively high density of final states gives rise to a momentum relaxation time of the order of a few femtoseconds \cite{Gusev1998}, whereas the very low density of states in semimetals suppresses relaxation towards thermodynamic equilibrium, particularly in equilibration between the electron and hole distributions\cite{Blount1959}.
Bismuth is an indirect semimetal---its electron and hole band extrema lie in different regions of the Brillouin zone ($L$ and $T$ points respectively) --- so that electron-phonon scattering is the dominant mechanism by which equilibrium is established between electrons and holes~\cite{LopezPR1968}.
In semiconductors, whereas thermalization of carriers within each band occurs on a sub-picosecond timescale, an energy band gap separates the electron and hole state energies, eliminating recombination by electron-phonon interaction and giving rise to typical recombination times of the order of ns.

In the group V semimetals, the photoexcitation of modest plasma density leads to large amplitude atomic motion corresponding to a coherent zone-center fully symmetric (A$_{1g}$) optical phonon \cite{Cheng1991,stevens2002}. The initial photoexcited electronic distribution is determined by the intensity, photon energy and polarization of the incident light. The intensity determines the overall number of carriers, the photon energy their energy distribution within the bands and the optical polarization gives rise to anisotropy in the distribution (i.e. differences in the photoexcited carrier occupation) between equivalent regions of the Brillouin zone. This distribution relaxes back to the thermal equilibrium distribution over time in a series of stages. Electron momentum relaxation (due to electron-electron and electron-phonon scattering) removes the anisotropy on a timescale of the momentum scattering time. The electronic energy distribution relaxes more slowly because many scattering events are required to change significantly the energy of an electron. When this process is due to electron-phonon scattering, heat is simultaneously transferred from the electronic to the vibrational degrees of freedom, raising the lattice temperature. Finally, scattering between conduction and valence bands, which can occur by electron-phonon scattering only when electrons are very near the bottom of the conduction band, establishes equilibrium in the occupation of conduction and valence bands on the timescale of the electron-hole recombination time\cite{Blount1959}.

Recently there has been  considerable interest in the dynamics of the A$_{1g}$ mode in the limit of a dense electron-hole plasma, particularly for bismuth \cite{deCamp2001PRB,hase2002, Sokolowski-Tinten2003,Hase2004,fahy2004,Murray2005,Fritz2007,Beaud2007,Johnson2008}.  \citet{Hase2004} observed that the frequency of the mode is impulsively softened and chirped  toward the equilibrium value over a few picoseconds.
\citet{Murray2005} demonstrated both experimentally using double pulse excitation and theoretically using constrained density functional theory (DFT) and frozen phonon calculations, that the chirp was dominated by electronic softening and the evolution of the dense electron-hole plasma.
Later, the theoretical studies were extended to include softening of the entire phonon spectrum \cite{Murray2007}.
The general results for the $A_{1g}$ mode were confirmed by \citet{Fritz2007} using femtosecond x-ray diffraction to measure the quasi-equilibrium position and curvature of the excited state potential.
However, the results of measurement of ultrafast oscillations in the Debye-Waller factor by \citet{Johnson2009} suggested that the coupling of the plasma with low-frequency acoustic phonon modes may not be well described in current theoretical models\cite{Zijlstra2010,Johnson2010}.

The calculations in Ref. \onlinecite{Murray2005} and \onlinecite{Murray2007} assume that a single electron-hole pair is created by each absorbed photon and that, whereas intraband scattering rapidly establishes a Fermi-Dirac distribution of carriers in each band, the electron-hole recombination time is substantially longer than the period of the relevant phonon and, as a result, the chemical potential for the conduction bands differs from that of the valence bands (two chemical potential model).
This ``two chemical potential'' model is implicit in the previous analyses of low-temperature ultrasonic attenuation in semimetals \cite{LopezPR1968} (see also Ref.~\onlinecite{Blount1959}).
In contrast, \citet{zijlstra2006} performed frozen phonon calculations on the zone-center mode, also assuming rapid scattering and equilibration throughout all conduction and valence bands, but using a single chemical potential, in which the carrier density is determined solely by the temperature of the electron-hole plasma.
\citet{Johnson2008} performed depth resolved femtosecond  x-ray diffraction to measure the phonon dynamics and concluded that the two-chemical potential model may be appropriate only on a time-scale less than a single phonon period, transitioning a single chemical potential model. \citet{Giret2011} have analyzed the Bragg peak intensities, reported in {Ref.\ \onlinecite{Fritz2007}}, using the model that assumes rapid thermalization of the carrier distribution throughout all bands and finds agreement with the experimental results that is as satisfactory as that obtained with the two chemical potential model.
Recently \citet{Sciaini:2009qc}  measured a 2--3 ps lattice thermalization time of the photoexcited carriers in the limit of dense excitation using ultrafast electron diffraction, indicating a rapid energy transfer from carriers to the lattice.
Still the dynamics of the photoexcited carriers   are not well understood.
The rapid transfer of energy to the lattice suggested in Ref. \onlinecite{Sciaini:2009qc} may call into question the detailed quantitative conclusions of analyses (including those of Refs.~\onlinecite{Murray2005,Fritz2007,Murray2007,Johnson2009,zijlstra2006,Giret2011}) that have assumed insignificant heat transfer from the electron-hole plasma to the lattice vibration modes in the first few picoseconds following photo-excitation.

Here we report results of ultrafast counter-propagating pump-probe measurements of the photoexcited carrier dynamics in a series of optically thick single crystal bismuth films of various thickness at room temperature.
By measuring the transient reflectivity signal on the opposite face of the film, we are able to separate the effects of carrier dynamics, lattice heating and strain, whereas the effects are coupled when probed on the same face as the pump \cite{boschetto2008PRL}.
In this manner, we are able to measure separately the ambipolar diffusivity along the trigonal axis and the carrier recombination rate.
By comparing the reflectivity changes at the front and back of the film at delay times from 100 to 500 ps, we find that the film is heated much more at the front than at the back, consistent with recent grazing incidence x-ray diffraction measurements on the thermal transport in thin Bi films\cite{Sheu2011SSC,walko2011JAP} and confirming unambigously that most of the heat transfer from the electron-hole plasma to the lattice occurs in the 5 ps  following photoexcitation, before the plasma diffuses to a uniform density within the film  \cite{Sciaini:2009qc}.

The principal aim of the current experiment is to estimate the timescales of momentum relaxation, heat transfer to the lattice, and electron-hole recombination. In particular, by investigating the diffusion of carriers to the back of films of various thickness, we estimate the momentum scattering time. By comparing the final heating of the front and back sides of the film, we put lower limits on the rate of heat transfer to the lattice, and by observing the decay of the plasma at the back of the film, we estimate the timescale of electron-hole recombination. We interpret each of these timescales within a common framework of electron-phonon scattering and with reference to the known band structure of bismuth and find qualitative agreement between the various relaxation times observed.
We discuss the relationship of these results to early studies~\cite{LopezPR1968} of low-temperature electron-hole recombination rates and to the relevant electronic density of states in various regimes of electronic excitation.
We argue, based on a picture of electron-phonon scattering, that the relaxation of carrier momentum within the hole and electron bands of the initially excited plasma should occur at a rate typical of metals (with a relaxation time of the order of a  several femtoseconds), that cooling of the plasma and corresponding lattice heating should occur on a timescale of a few picoseconds and that electron-hole recombination should occur substantially more slowly than either of these processes. We argue that as the plasma cools towards room temperature the electron-hole recombination rate and the plasma cooling rate should be of similar  magnitude as a greater proportion of the excited carriers lie in the energy range where the conduction and valence bands overlap.

The rest of this paper is organized as follows. In Section \ref{sec:exptl_details} we give the details of our experimental setup. The results of experiments with low optical pump fluence are presented and discussed in Section \ref{sec:low_fluence}. Measurements at high optical pump fluence are discussed separately in Section \ref{sec:high_fluence}. In Section \ref{sec:bands} we discuss the band structure of bismuth and the role of electron-phonon scattering in the various momentum, energy and electron-hole relaxation processes. In the concluding Section \ref{sec:conclusions}, we summarize the experimental findings and the principal conclusions of this study. Some technical details of electron-phonon scattering and carrier relaxation are presented in the Appendix.

\section{Experimental details} \label{sec:exptl_details}

Various thickness single crystal c-axis Bi films were grown on $1\times1$ cm$^{2}$  sapphire substrates (c-axis $\sim 0.5$ mm thick) by molecular beam epitaxy (MBE).
We conduct experiments using Ti:sapphire lasers; pulses of 85 fs duration, energy 3 nJ/pulse and wavelength centered at 830 nm were produced  at a repetition rate of ~85 MHz (low density excitation) and pulses of 100 fs duration, energy 4 $\mu$J/pulse, and wavelength centered at 800 nm were produced at a 250 kHz repetition rate (high density excitation).  The laser pulses are short compared to the $A_{1g}$ phonon frequency, 2.92 THz  at room temperature.
Approximately 5 $\%$ of the pulse energy is used for the probe beam. A mechanical delay stage controls the pump-probe delay.
The pump and probe beams are focused by lenses with 30 cm and 15 cm focal length respectively and have a diameter of $\sim$1 mm before the lenses.
The data are reproducible to better than 10$\%$ in the relative reflectivity, $\Delta$R/R, to  $\sim$10$^{-6}$ using a mechanical chopper, balanced detector and lock-in amplifier.

\section{Results and discussion for low excitation}\label{sec:low_fluence}

\subsection{Surface pump and surface probe}

\begin{figure}[tb]
\begin{center}
\includegraphics[width=3.2in]{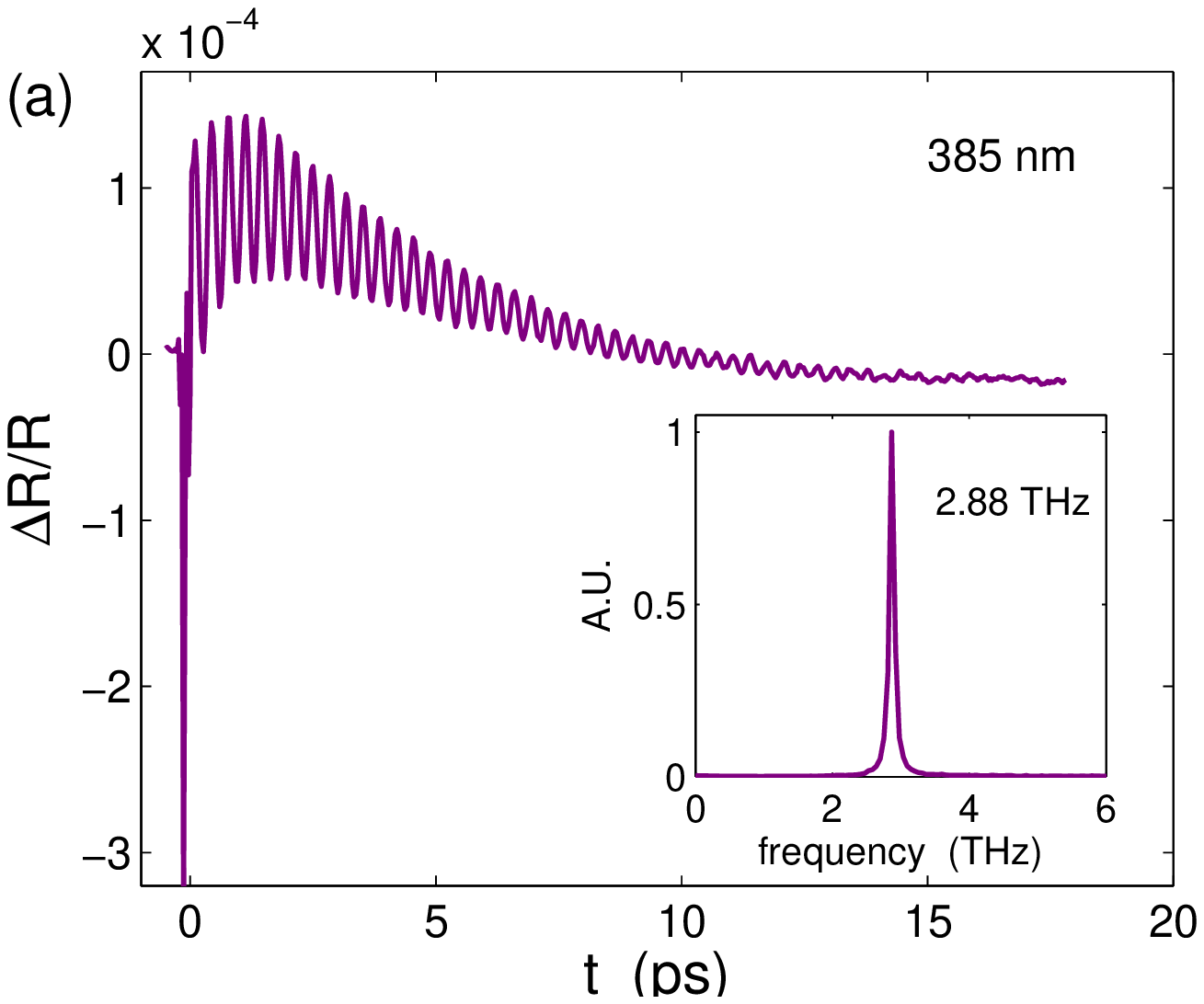}
\includegraphics[width=3.2in]{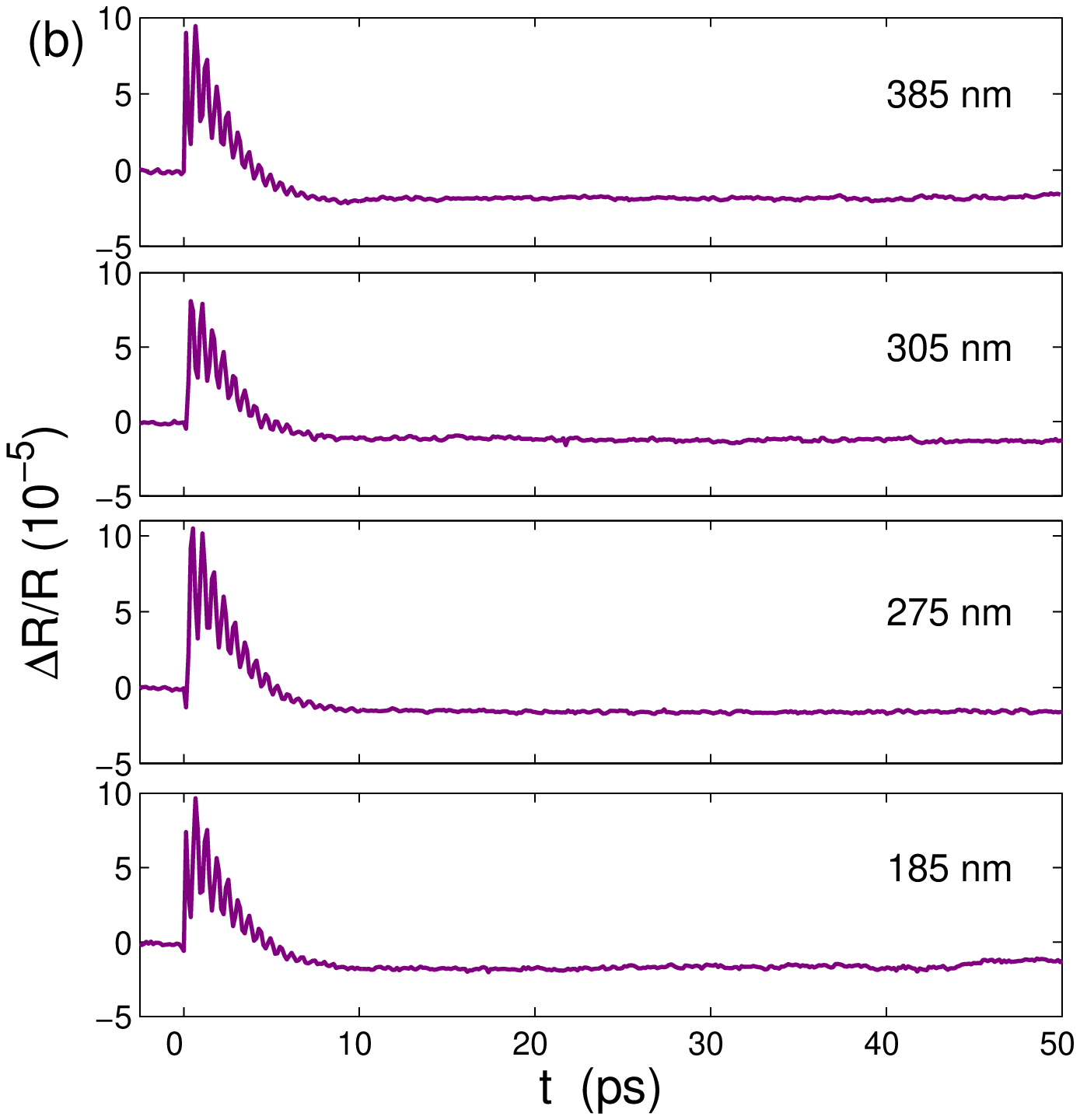}
\caption{ \label{f:SpumpSprobe} (color online)  (a) A$_{1g}$ phonon oscillation. The inset shows the Fourier transform of A$_{1g}$ oscillation. (b) Conventional surface-pump/surface-probe data. }
\end{center}
\end{figure}

Fig. \ref{f:SpumpSprobe} shows ultrafast data for  conventional surface (air/Bi) pump and surface probe  for four different thickness films, ranging from 185--385 nm.  We estimate the absorbed fluence to be about 0.8 $\mu$J/cm$^2$.
The reflectivity data shows oscillations at 2.88 THz, near the room temperature  $A_{1g}$ frequency,  demonstrating that we are in the low fluence regime where the phonon frequency is not substantially softened by the photoexcitation. Phenomenologically,  the relative reflectivity change is
\begin{equation}
\frac{\Delta R(t)}{R}=\frac{\partial \ln R}{\partial x}\Delta x+
\frac{\partial \ln R}{\partial n}n+\frac{\partial \ln R}{\partial \eta}\eta+\frac{\partial
\ln R}{\partial T} \Delta T, \label{e:dR}
\end{equation}
to first order in the $A_{1g}$ phonon coordinate $x$, photoexcited carrier density $n$, strain $\eta$, and lattice temperature  $T$, all of which are functions of time $t$ since laser-excitation.
Generally the partial derivatives are not known; however, we can deduce their signs from the data as discussed below:  $\partial R/\partial x > 0$ ($x$ driven towards the symmetric non-Peierls distorted structure), $\partial R/\partial n>0$, $ \partial R/\partial \eta < 0$ for tensile and $>0$ for compressive strain, and $\partial R/\partial T<0$.
We note that, in expanding the reflectivity as in Eq.~(\ref{e:dR}), we do not presuppose any particular constraint on the carrier density $n$ --- for example, under certain physical conditions, $n$ could be determined by the electronic temperature, $T_p$. We assume second order and higher terms are negligible.

The difference between the surface probe data (Fig. \ref{f:SpumpSprobe}(b)) for various film thicknesses is small.
This is to be expected; because of strong absorption, carriers are generated near the surface ($\sim$ 15 nm laser penetration depth) and the probe beam only detects the reflectivity change in this region.
We see  $\Delta R(t)/R$ first increasing rapidly due to the sudden increase in carrier density and then slowly decreasing as the carriers relax via multiple processes.
As discussed above, the oscillations at early times are due to the A$_{1g}$  phonon ( clearly resolved in Fig. \ref{f:SpumpSprobe}(a)  but aliased by the relatively coarse sampling in Fig. \ref{f:SpumpSprobe}(b)).
The overall reduction of the reflectivity signal during the first few picoseconds is due to a competition of effects from carrier relaxation through ambipolar diffusion, recombination, lattice heating and the propagation of an acoustic pulse generated through a combination of rapid thermal expansion and the acoustic deformation potential interaction. The signal becomes negative after approximately 10 ps as the strain pulse leaves behind a thermally expanded region at the surface, which decays on the much slower time scale of heat conduction in the lattice.
We note that an optically thin sample would mitigate the effects of the inhomogeneous carrier distribution; however it would not decouple the effects of the coherent phonon, recombination, lattice heating or strain  on the reflectivity change since they occur on overlapping time scales.

\subsection{Surface pump and interface probe}

\begin{figure}[tb]
\begin{center}
\includegraphics[width=3.6in,height=3.6in]{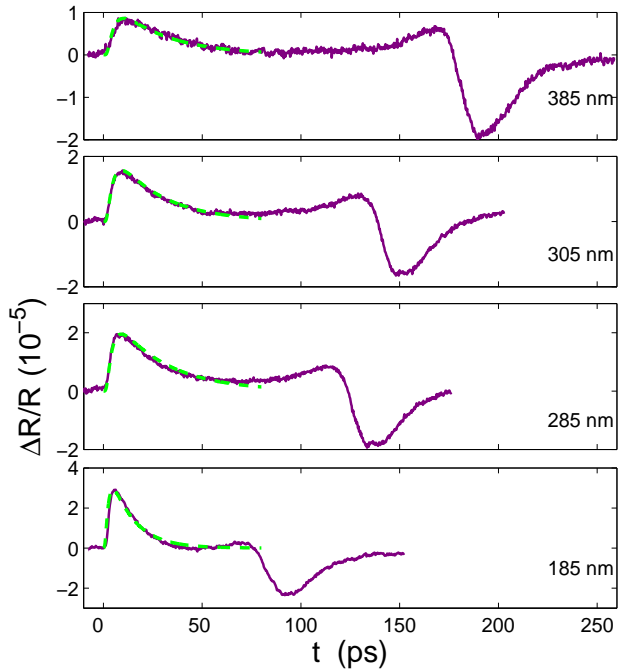}
\caption{ \label{f:SpumpIprobe} (color online)   Counter-propagating surface-pump/interface-probe data (solid line) and fits for carrier signals (dash line from 0--80 ps), described in the text. Each set of data is collected up to the time when the first acoustic strain is observed.}
\end{center}
\end{figure}

\begin{figure}[tb]
\begin{center}
\includegraphics[width=3.2in]{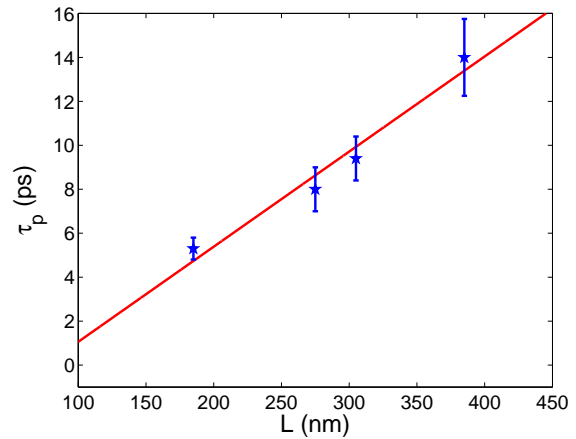}
\caption{ \label{f:slope} (color online) The peak time,  $t_p$,  for carriers to arrive at the interface of the film  as a function of film thickness, $L$, showing an approximately linear relationship with a slope on order 30 ps/$\mu$m.}
\end{center}
\end{figure}

Fig. \ref{f:SpumpIprobe} shows the relative reflectivity change for the probe incident on the
Bi/sapphire interface.
Clear differences are seen for various film thickness, compared to the surface-probe data.
There is no coherent A$_{1g}$ phonon signal present on the interface-probe data, consistent with a localized excitation of the zero group velocity mode in the laser-excited volume of the optically thick films \cite{A1gNote}.
Thus, to the extent that we can ignore heating due to recombination at the  interface, both the first and fourth terms on the right hand side of Eq. (\ref{e:dR}) are absent in the interface signal.
The bipolar feature at late times  ($> 50$~ps) is attributed to the acoustic strain pulse.
We use the arrival time of the acoustic pulse to calibrate the film thicknesses relative to the thinnest film (185 nm) which we measured using grazing incident x-ray reflectivity. The measured speed of sound  for the thinnest film, 2.15 $\pm$ 0.2 km/s, compares with 1.972 km/s of Ref. \onlinecite{SoundSpeed}.  The uncertainty arises from the difficulty in determining the precise pulse propagation time due to the finite temporal width of the acoustic pulse.  Well before the arrival of the acoustic pulse, the signal peaks at a time ($t_p$) that is approximately linearly proportional to the film thickness (see Fig. \ref{f:slope}). This peak is attributed to carriers that have diffused across the film before recombining.  Note that the carriers' signal is well separated from the acoustic pulse which takes approximately $15 t_p$ to arrive at the interface.

The shape of the early-time component (carrier peak) of the reflectivity signal depends on both ambipolar diffusion and recombination. To model the dynamics, we assume that carrier diffusion is one-dimensional (in the coordinate $z$ along the c-axis).  This is appropriate since the laser pump diameter, $\sim$150 $\mu$m, is many orders of magnitude greater than the film thickness.  In the limit that the recombination time $\tau$ and the ambipolar diffusivity $D$ are independent of carrier density, $n$,
\begin{eqnarray}
\frac{\text{d}n}{\text{d} t} = D \frac{\text{d}^2 n}{\text{d} z^2}- \frac{n}{\tau}.
\label{e:n}
\end{eqnarray}
It is instructive to consider the limiting case of an infinitely thick
sample with a $\delta$-function excitation at $t=0$, for which

\begin{equation}
n(z,t) = \frac{N}{\sqrt{\pi D t}}\text{e}^{-\frac{z^{2}}{4Dt}}\text{e}^{-t/\tau},
\end{equation}
and we find that the maximum density at a given depth $z$ occurs at time,
\begin{equation}
t_{p} = \frac{\tau}{4} \left( \sqrt{1+\frac{4z^{2}}{D\tau} } -1\right). \label{e:tp}
\end{equation}
Note that deep enough into the bulk $t_p$ is linearly proportional to $z$ and
\begin{equation}
\frac{\text{d}t_{p}}{\text{d} z} =   \frac{1}{2}\sqrt{\frac{\tau}{D}},~z \gg \sqrt{D\tau/ 4}.  \label{e:approxtp}
\end{equation}
This approximation is reasonably accurate for the interface $z=L$ of a
relatively thick film. Thus, the linear dependence seen in Fig. \ref{f:slope} suggests that our samples satisfy $L\gg\sqrt{D\tau/4}$.
Moreover,  Eq. (\ref{e:approxtp}) allows us to make a rough estimate for $D\approx$ 70 cm$^{2}/$s given the $\sim$30 ps/$\mu$m slope in Fig. \ref{f:slope} and $\sim$25 ps decay from the reflectivity signal in the time period 10--50 ps after the pump.

\begin{figure}[tb]
\begin{center}
\includegraphics[width=3.2in]{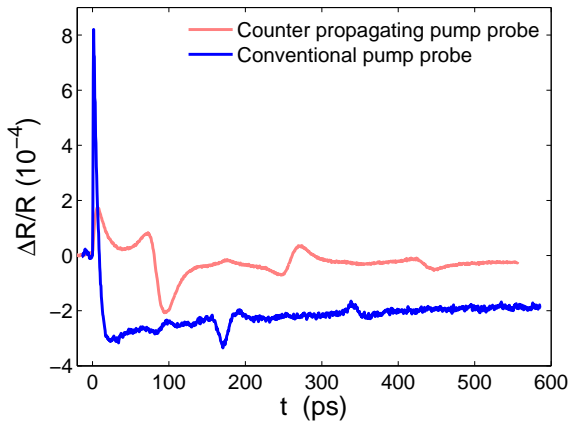}
\caption{ \label{f:Frontback} (color online) Comparison of 185 nm film between surface-probe and interface-probe data both pumped from Bi surface at {$n_0\sim$ 10$^{20}$ cm$^{-3}$} assuming linear absorption. Multiple acoustic pulses and echoes are shown.  }
\end{center}
\end{figure}

\begin{table}
\begin{center}
\begin{tabular}{|c|c|c|}
  \hline
      L &D (cm$^2$/s) &$\tau$ (ps) \\
   \hline
  185 nm &18$\pm$5 & 12$\pm$2\\
  275 nm & 24$\pm$6 & 26$\pm$4 \\
   305 nm & 28$\pm$5 & 24$\pm$3 \\
   385 nm & 40$\pm$14 & 26$\pm$6\\
  \hline
\end{tabular}
\caption{Fit results from various films.}
\label{t:fits}
\end{center}
\end{table}

\begin{figure}[tb]
\begin{center}
\includegraphics[width=3.2in]{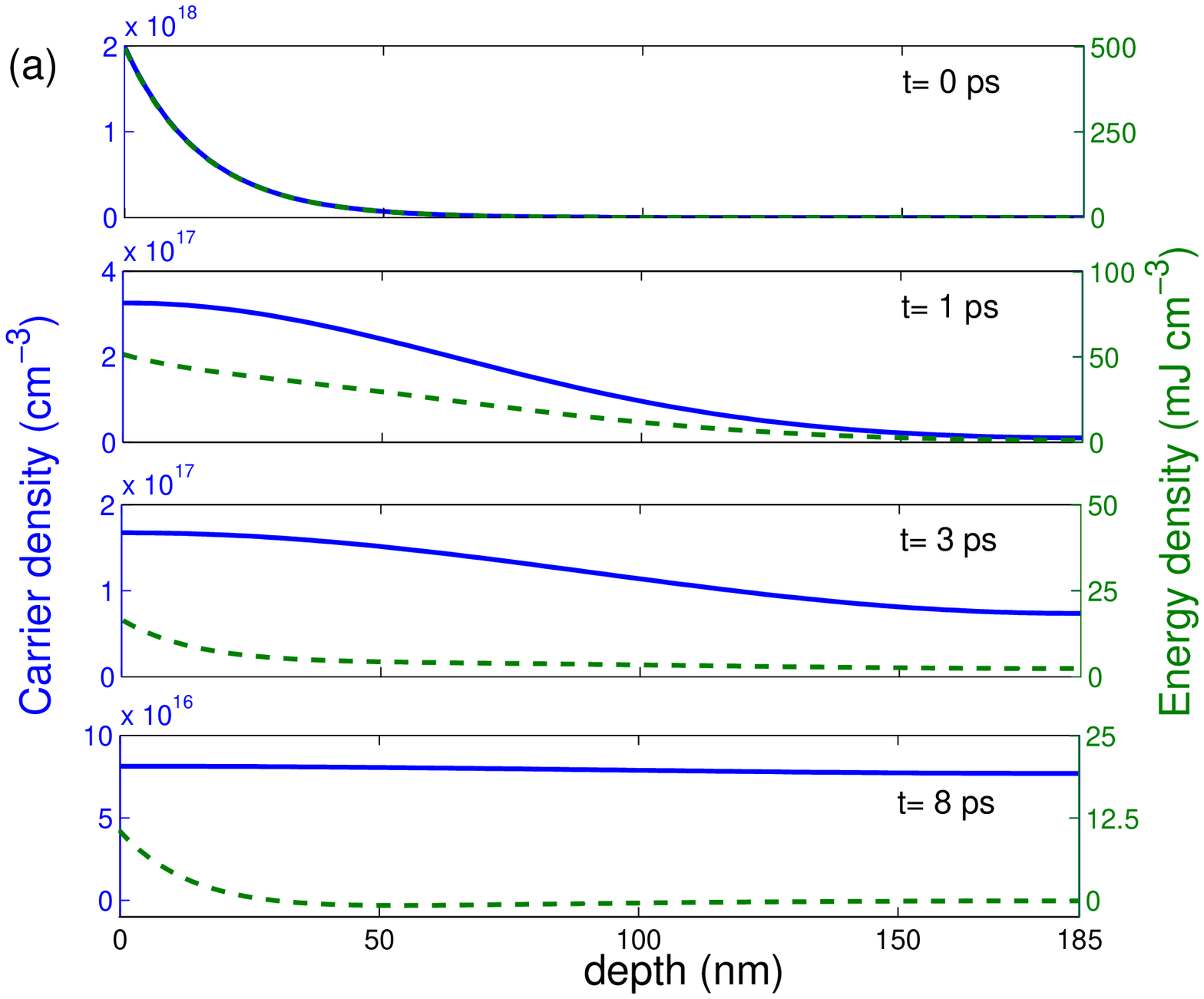}
\includegraphics[width=3.2in]{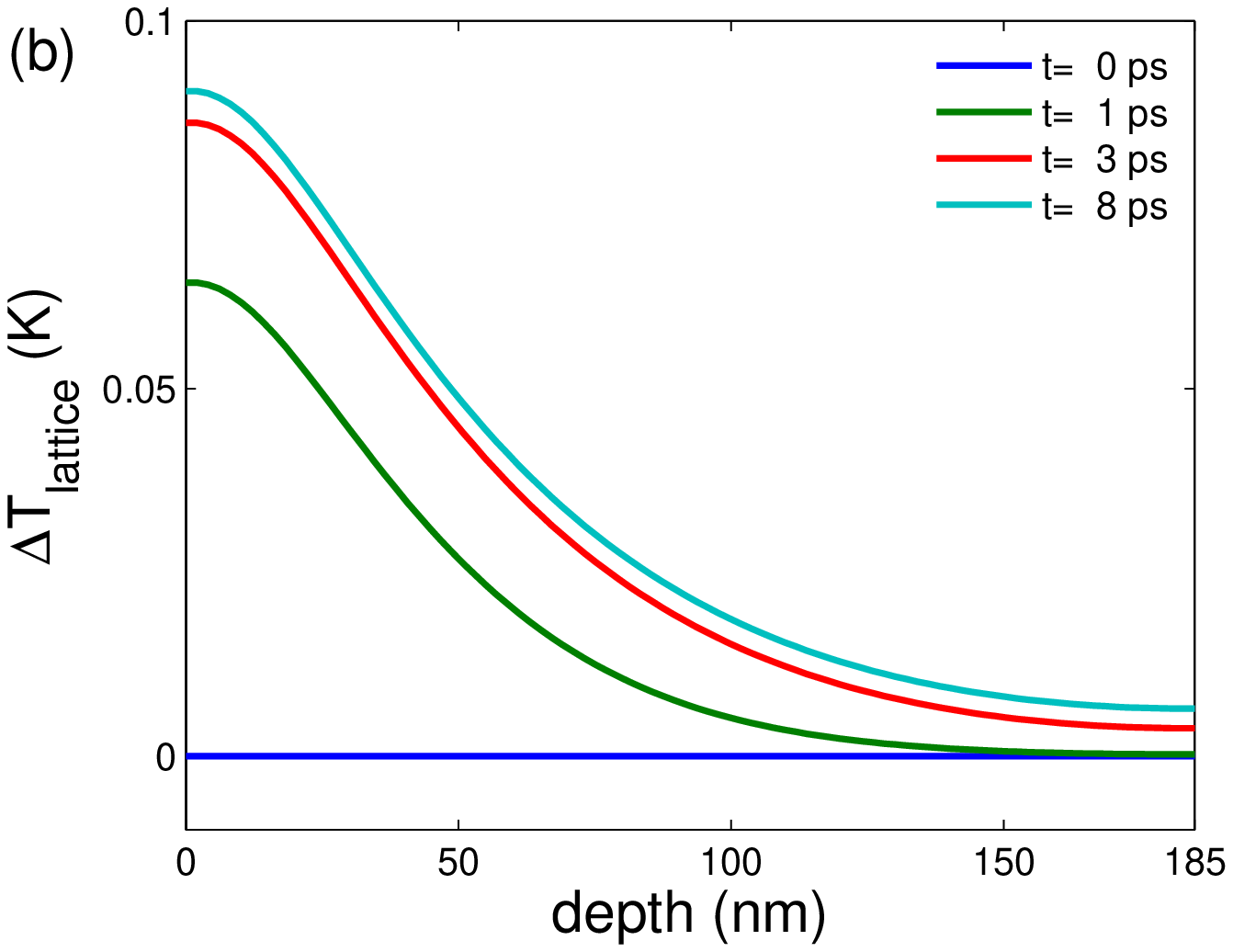}
\caption{ \label{f:EnergyDensity} (color online) Simulation for 185 nm film using diffusion constant 18 cm$^2$/s, recombination time 12 ps, and lattice thermalization time 1.5 ps for (a) carrier density (solid line with left axis) and energy density (dash line with right axis) and for (b) the increased lattice temperature as a function of depth at various time delays with initial carrier density $n_0$=2$\times$10$^{18}$ cm$^{-3}$. }
\end{center}
\end{figure}

To better model the data (and allow for variations in the individual samples),
we solve Eq. (\ref{e:n}) numerically for the finite crystal case, treating $D$ and $\tau$ as fitting parameters.
 We assume that the carriers are confined at the surface ($z=0$) and bismuth-sapphire interface ($z=L$),
$\text{d}n/\text{d}z |_{z=0, \text{L}} = 0$,
which is valid so long as surface recombination is slow.
The initial photoexcited carrier distribution is assumed to decay exponentially over a 15 nm laser penetration depth with a single electron-hole pair per incident photon.
The initial carrier density $n_0$ for the data in Fig. \ref{f:SpumpIprobe} is estimated to be about 2$\times$10$^{18}$ cm$^{-3}$ at the surface.
The probe is distributed across the laser penetration depth.
The overall amplitude is normalized to the individual data of various films.
The  results of the fits are summarized in Table \ref{t:fits}.

 From the ambipolar diffusivity, $D$, we can make a rough estimate of the momentum relaxation in the relaxation time approximation. Assuming an average band effective mass $m^*$ of a free electron, and taking the relaxation time as
$\tau \approx m^*D/(kT_p)$, we obtain an estimate of the scattering time, equal to 0.6~ps for a room temperature plasma. In the initial hot photo-excited plasma, for which $kT_p \approx 0.5 $~eV, the relaxation time reduces to approximately 30 fs.  This is reasonably consistent with the results based on density of states arguments in Section \ref{sec:bands}.

We note that the height of the first peak in $\Delta R/R$ for the low-fluence data indicates that the rate of carrier recombination prior to $t_p$ is not substantially faster than in the time after $t_p$. If it were, a substantially smaller peak in $\Delta R/R$ would be observed. As we will see later, this is not the case for high-fluence data.

In Fig. \ref{f:Frontback}, the surface pump-probe signal drops below zero after about 10-20~ps, with a slow recovery on the nanosecond time-scale.
This initial drop is caused by acoustic strain propagation from the surface through the optical absorption depth, leaving a thermally expanded lattice in its wake, while the subsequent very slow decay is attributed to cooling due to heat conduction.
We note that the background reflectivity change due to this thermal expansion is much greater at the surface of the film than the corresponding background at the interface.
Therefore the front surface of the film has been heated much more than the back interface.  This is consistent with recent time-resolved x-ray diffraction experiments on the thermal transport in thin Bi, in which the thermal gradient is maintained for a time on order nanoseconds for comparable film thicknesses\cite{Sheu2011SSC,walko2011JAP} .
This demonstrates that lattice heating must occur substantially faster than the time ($\approx 5$~ps) taken for diffusion to produce a uniform plasma in the 185~nm thick film.
Thus, the carriers cool by heat transfer to the lattice in a time substantially less than 5~ps and the carriers reaching the back surface of the film must be relatively cold, compared to the initial photo-excited plasma.
This is  also consistent with electron diffraction measurements of thin films reported in Ref.~\onlinecite{Sciaini:2009qc} that suggest a lattice heating time of $2-2.5$~ps.
Although the plasma cools in substantially less than 5~ps, its density decays with a much longer time constant (10-30~ps).

This indicates that the decay of the plasma density is determined by the electron-hole recombination rate and we therefore estimate the recombination time to be of the order of 20~ps.
(However, bearing in mind the discussion of cooling of the hot plasma, below, we cannot  entirely rule out the possibility that the cooling rate of the plasma decreases as it cools towards the lattice temperature.)
Using separate decay rates for carrier recombination and lattice thermalization:

\begin{eqnarray}
\frac{\text{d}n(z,t)}{\text{d} t} =  - \frac{n(z,t)}{\tau_{e-h}} + D \frac{\text{d}^2 n}{\text{d} z^2},\\
\frac{\text{d}E(t)}{\text{d} t}=- \frac{E(t)}{\tau_Q},\\
C_v\frac{\text{d}\Delta T_{l}(z,t)}{\text{d} t}= \hbar\omega e^{-t/\tau_{Q}}(\frac{1}{\tau_{e-h}}+\frac{1}{\tau_Q})n(z,t),
\label{e:2decay}
\end{eqnarray}
where $\tau_{e-h}$ is recombination time, $E(t=0)=\hbar\omega$ is photon energy absorbed per electron-hole pair with decay time $\tau_Q$ after excitation, $C_v$ is lattice specific heat, and $T_l$ is lattice temperature,  we show carrier density, energy density and lattice temperature change (without taking thermal diffusion into account) in Fig. \ref{f:EnergyDensity} for the 185 nm. The hot plasma ($t=0$) cools much faster than the electrons and holes recombine.

The rate of temperature loss in the semimetal plasma is comparable to the
typical rates observed in metals (a few picosecond or less). However, the rate of
electron-hole recombination (or, equivalently, scattering between the $T$
and $L$ valleys) is substantially slower.
Our values of the recombination time, as observed by the decay of the plasma at
the back side of the film during the interval 10-50 ps following
photo-excitation, are  considerably larger than a simple extrapolation of the fit of
low-temperature acousto-magneto-optical effect data, given in \citet{LopezPR1968}.

\section{Results and discussion for High excitation}\label{sec:high_fluence}
\begin{figure}[tb]
\begin{center}
\includegraphics[width=3.2in]{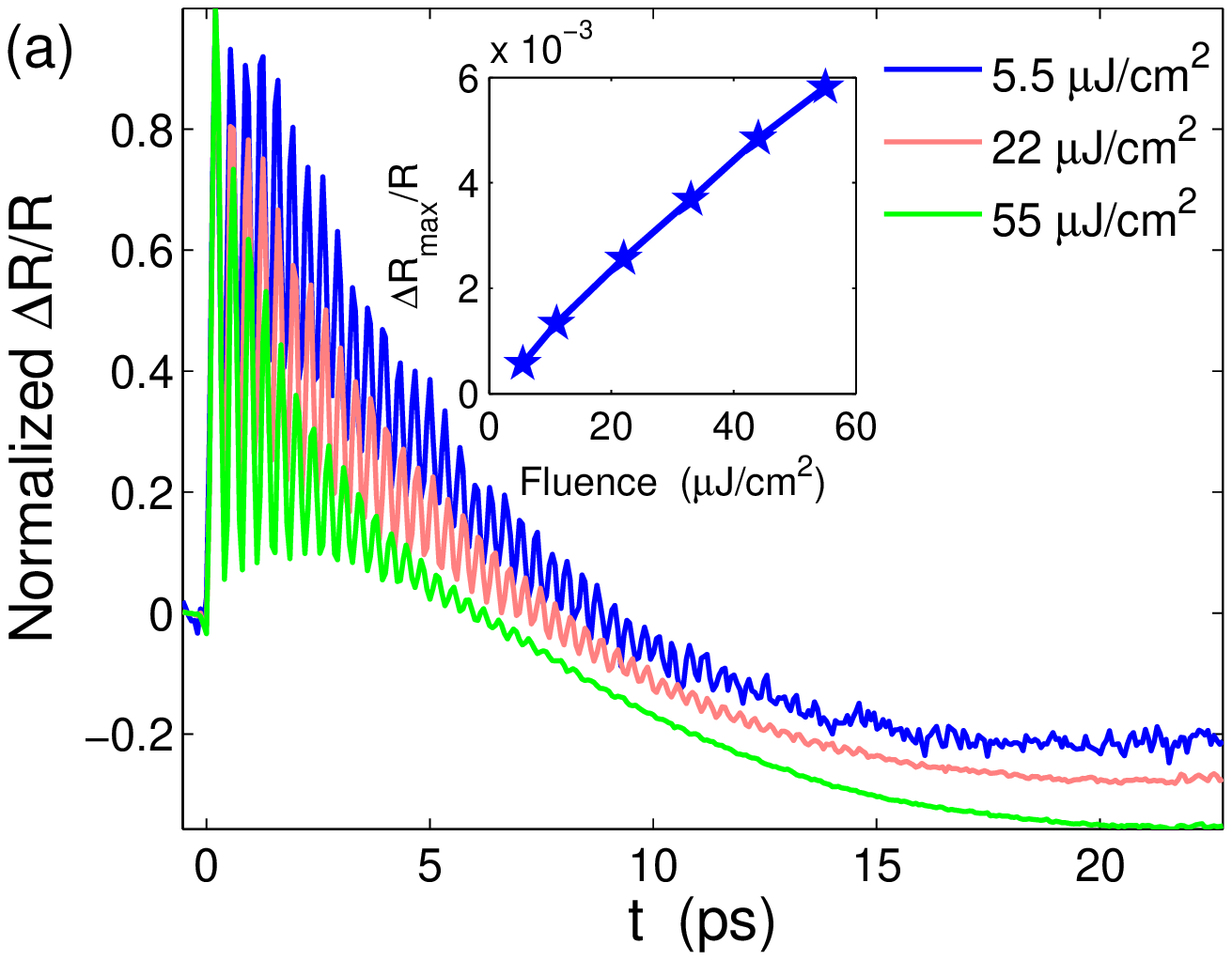}
\includegraphics[width=3.2in]{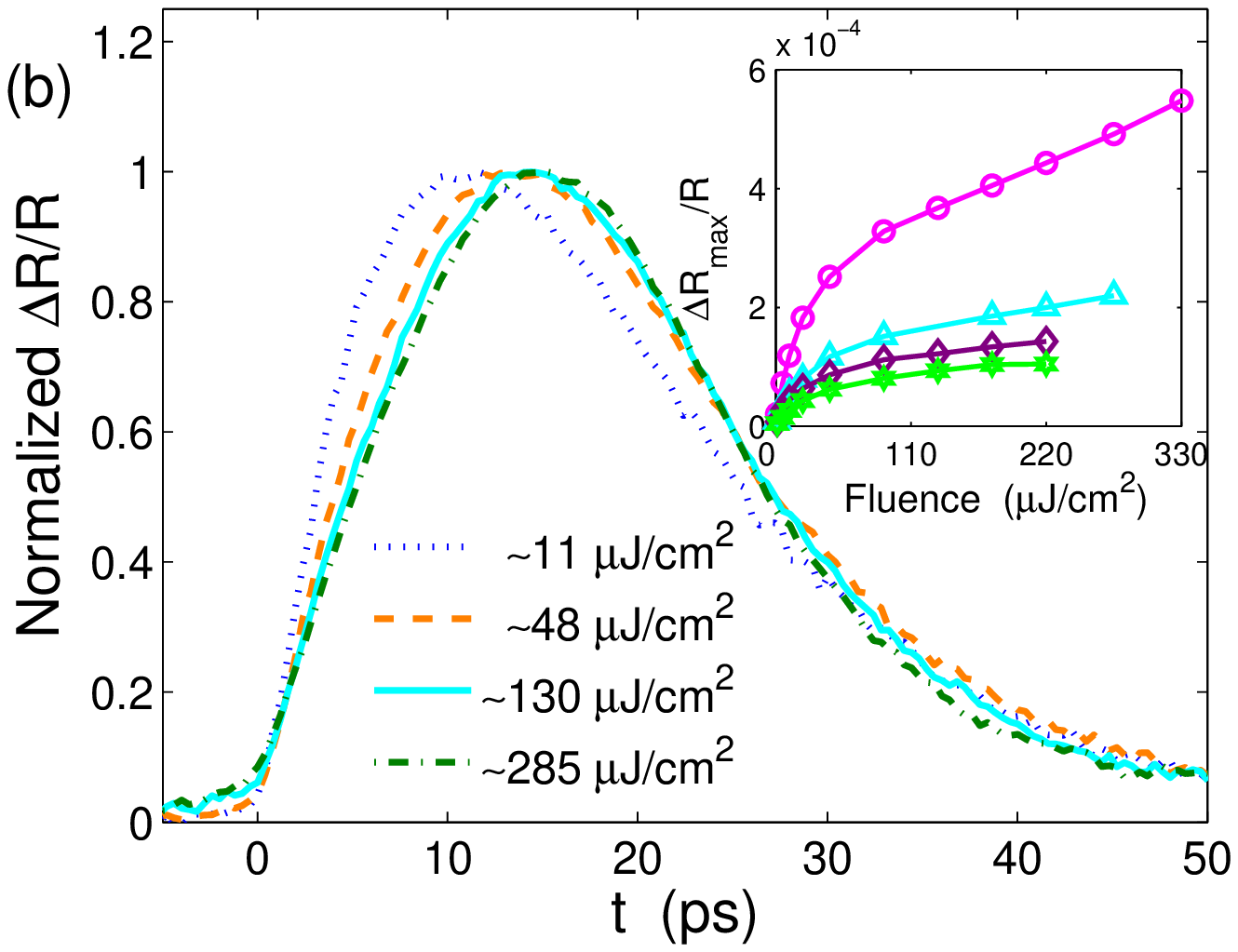}
\caption{ \label{f:power} (color online) (a)  Surface-pump/surface-probe data for 185 ~nm film at various absorbed fluence. The inset shows the maximum intensity at different fluence, which is similar for various film thickness.  (b) Surface-pump/interface-probe data for 385 ~nm film at various absorbed fluence. $n_0$ $\sim$ 3$\times$ 10$^{19}$ cm$^{-3}$ is estimated from absorbed fluence of $\sim$11 $\mu$J/cm$^2$ assuming linear absorption. The peak time is delayed for higher fluence.  The inset displays the peak intensity for various film thickness at different fluence. From lower to upper lines are 385, 305, 275 and 185 nm respectively.}
\end{center}
\end{figure}

For the carrier density $\gtrsim 10^{20}$ cm$^{-3}$, the dynamics become more complex. Fig. \ref{f:power} shows data as a function of time at a range of pump laser fluence for (a) surface pump and probe of 185 nm film and (b) surface pump interface probe of 385 nm film , scaled in peak amplitude for comparison. The individual inset shows how the amplitude varies with fluence for (a) 185 nm film and (b) the different thickness films.  The peak intensity of the carrier signal in the surface-pump/interface-probe data, shown in Fig. \ref{f:power}(b) , saturates and the peak time $t_p$ becomes longer and carrier recombination becomes faster.   On the other hand, for the surface pump and probe, the maximum (carrier density) and minimum (thermal) component of the relative reflectivity remain linear up to much higher fluence.   We were not able to determine the level at which this signal saturates due to potential artifacts at the highest fluence for the surface probe.  Nonetheless, it appears that at high densities the recombination and diffusion become nonlinear.  The increased plasma decay rate in the initial dense plasma reduces the amplitude $\Delta R_{n(max)}/R$ of the carrier peak of the counter-propagating reflectivity at $t=t_p$ by up to a factor of 5 at the highest fluences, indicating that the plasma density decays at an average rate of 0.3 ~ps$^{-1}$ at early times ($t<5$~ps), comparable to the rate of heat transfer to the lattice.
There are a few possibilities for the increased recombination in the  non-linear regime: (a) the Auger channel begins to dominate over the phonon recombination channel at high enough density (when $n > \sqrt{\gamma/A}$, where $\gamma$ is the linear recombination rate and $A$ is the coefficient for Auger recombination); (b)  the phonon-assisted recombination rate is enhanced due to the presence of more thermally excited phonons because the lattice is itself heated; (c) the opening of new recombination channels because the electronic structure is substantially altered when the Peierls distortion is partially reversed in the photo-excited material --- for example, if the band overlap increased, this would open up a higher density of states for the indirect phonon-assisted recombination, or if the direct gaps either at $T$ or $L$ were reduced enough, we could have phonon-assisted recombination across the direct gap (see Section \ref{sec:bands} below).
At this point, the nature of the nonlinearity cannot be determined without further experiments and more sophisticated analysis.

\section{Discussion --- Band structure and electron-hole recombination in Bismuth}\label{sec:bands}

\begin{figure}
\begin{center}
\includegraphics[width=3.3in]{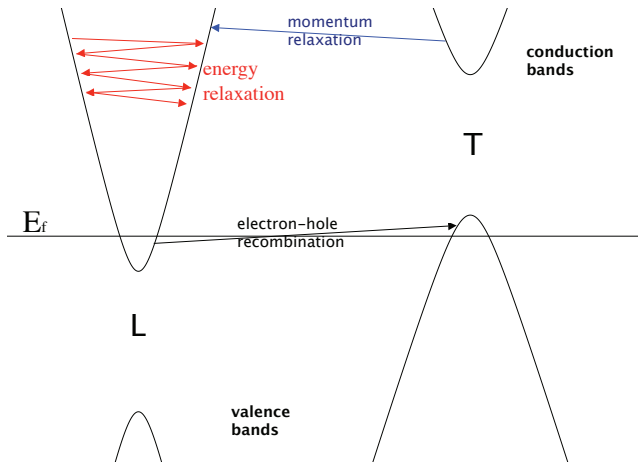}
\caption{ \label{f:BiBand} Schematic of the Bi band structure near the Fermi energy, showing typical electron-phonon scattering processes.  There is a small, $\sim$38 meV  overlap between the valence band maxima at the $T$ point (holes) and the conduction band minima at the $L$ points (electrons), such that carrier recombination occurs over a small region of phase space involving the emission or absorption of a  phonon with momentum near $X$.}
\end{center}
\end{figure}

Bismuth has two atoms in the primitive cell and 5 valence electrons per atom,
giving 10 valence electrons per primitive cell, which almost completely fill the lowest five valence bands.
However, due to a 38 meV band overlap between the fifth and sixth bands, electrons occupy the sixth band near the three $L$-points in the Brillouin zone, with a carrier density $\sim10^{17}$ cm$^{-3}$ ($\sim 2 \times 10^{-5}$ per unit cell) at low temperature \cite{LopezPR1968} (see also Ref.~\onlinecite{Isaacson1969}).
A corresponding number of holes occur near the $T$-point in the fifth band.
The Fermi level lies approximately $26$~meV from the conduction band minimum and $12$~meV from the hole band maximum.
The electronic density of states per unit cell is approximately $2 \times 10^{-3}$~ eV$^{-1}$ for electrons and $3 \times 10^{-3}$~ eV$^{-1}$ for holes at the Fermi energy. This gives an estimate of the typical density of states in the region of the overlap of the conduction and valence bands.
On the other hand, the typical density of states is approximately 1~eV$^{-1}$ at energies 0.5-1.0~eV from the Fermi level\cite{gonze1990}.
Fig.\ \ref{f:BiBand} shows schematically the electronic bands and scattering processes relevant for momentum relaxation, heat transfer to the lattice, and electron-hole recombination.

Phonon assisted electron-hole  recombination in Bi was discussed by Lopez in 1968. \cite{LopezPR1968} From the acoustomagnetoelectric effect,  Lopez measured an electron-hole recombination time $\sim$10$^{-8}$ s below 4 K,  which decreased rapidly above 6 K, assisted by the absorption or emission of phonons.
The acoustic attenuation is fitted with a combination of (temperature independent) defect scattering and two activated processes, with characteristic temperatures of $43$~K and $130$~K, which are consistent with recombination, assisted by two distinct phonons of energy $\hbar \omega$ equal to 3.7 meV and 11.2 meV, respectively, corresponding to acoustic and optical phonons at the $X$-point of the Brillouin zone.  Lopez considered the assignment of the component with activation temperature of $130$~K to electron-hole recombination assisted by the optical phonon at $X$ to be tentative and allowed for the possibility that this attenuation mechanism could be associated with hole scattering between valence band extrema at $T$ and elsewhere in the Brillouin zone.

Extrapolation to room temperature of the expression given in Eq.~26 of Lopez~\cite{LopezPR1968}, gives an electron-hole recombination time of approximately 0.5~ps, dominated by the 11.2 meV (130~K) phonon (if one neglects the component with activation temperature of 130~K, a room temperature recombination time of 6~ps is obtained).
However, although Lopez begins with a very general expression for carrier relaxation due to electron-phonon scattering in Section V.A of Ref.~\onlinecite{LopezPR1968}, the further derivation leading to Eq.~26 and the detailed analysis of the experimental data depend on the assumption that the system is at very low temperature and that the density of electronic states does not vary significantly over energies of a few $kT$.
Clearly this assumption breaks down for Bi at room temperature, where $kT$ is comparable to the valence-conduction band overlap, and is entirely invalid for a photo-excited electronic plasma at temperatures, where $kT \sim 0.5$~eV.
A quantitative prediction of the room temperature electron-hole recombination time from the results of Lopez is questionable, but values of the order of 0.5-10~ps are not unreasonable.

In Appendix A, we give a more general discussion of the role of electron-phonon scattering in establishing equilibrium of the chemical potentials of various bands (including electron-hole recombination as a particular case) and  in transferring heat from the electronic plasma to the lattice.
The phonon-assisted rate of relaxation of the chemical potential difference between two bands, $1$ and $2$, can be expressed qualitatively in terms of typical electron-phonon matrix elements and the average density of available final states, to which carriers can scatter by phonon absorption or emission:
\begin{equation}\label{rate_DOS}
\gamma_{1 \rightarrow 2} = {2 \pi \over \hbar} \langle{|M|^2}\rangle N(\hbar\omega) \langle D_{avail}\rangle/n
\end{equation}
where $\langle{|M|^2}\rangle$ is the average-square electron-phonon matrix element, which is expected to be of similar magnitude for short-wavelength acoustic modes and optical modes throughout the Brillouin zone,
$N(\hbar\omega) $ is the phonon occupation number, $\langle{D_{avail}}\rangle$ is the density of final electronic states per unit volume available at energies $\epsilon \pm \hbar \omega$, where $\epsilon$ is averaged over the energy of carriers in the initial band and $\omega$ is the typical optical phonon frequency, and $n$ is the carrier density.
The effective density of available states includes a blocking factor ($1-f$ for electrons and $f$ for holes, where $f$ is the occupation of the final state) arising from the Pauli exclusion principle.
Note that in our room temperature experiments all the bismuth phonon modes are in the classical regime, with $N(\hbar\omega) \approx kT_l/ \hbar\omega$, where $T_l$ is the lattice temperature.
The total carrier scattering rate $\gamma_p$ is a sum over the rates to all bands, including intraband scattering.
We show in Appendix A that the rate of the heat transfer to the lattice from a high-temperature electron-hole plasma gives a plasma cooling rate,
\begin{equation}\label{heat_relax_rate}
\gamma_Q
= \left[{\hbar \omega \over 3kT_p} \right] \left[ { \hbar\omega \over kT_l}\right]  \gamma_p,
\end{equation}
where $T_p$ is the plasma temperature, $T_l$ is the lattice (phonon) temperature and $\omega$ is the typical frequency for phonons involved in the electron-phonon scattering process.
We will discuss below the consequences of these results for electron-hole recombination and temperature relaxation in various regimes of photoexcitation.

\subsection{Momentum and energy relaxation of the hot plasma}

Immediately after photo-excitation, electrons and holes are created with energies approximately 0.75~eV into the conduction and valence bands, respectively. At these energies, the electronic density of states in each band is of the order of 1~eV$^{-1}$ and, based on Eq.~\ref{rate_DOS}, we expect a momentum relaxation rate $\gamma_p$ in this hot plasma approximately $10^{3}$ faster than the momentum relaxation rate in a room temperature plasma. Taking our crude estimate, above, for the carrier relaxation rate in the room temperature plasma, based on extrapolation of the low-temperature analysis of Lopez, where we expect a momentum relaxation time of the order of 0.5-10~ps, we would anticipate momentum relaxation time in the highly excited plasma to be of the order of several femtoseconds.
Indeed, this is in good agreement with the momentum relaxation time inferred in recent  experiments comparing spontaneous and impulsive stimulated Raman scattering in Bi and Sb, corresponding to the decay rate of the electronic forces driving the low-symmetry $E_g$ mode .   In these experiments, a decay time, $1/ \gamma_p \approx5-10$~fs is observed for the $E_g$ force at room temperature for Bi\cite{Li2011}.
On the other hand, taking a typical optical phonon energy in Bi to be approximately 10 meV, we can use Eq.~\ref{heat_relax_rate} to estimate a temperature decay time in the highly excited plasma to be
\begin{equation}
\tau_Q =  \left[{3 k T_p \over \hbar \omega } \right] \left[ { kT_l \over \hbar\omega }\right]
{1 \over \gamma_p } \approx 2-4~{\rm ps}.
\end{equation}
This estimate is in good agreement with inhomogeneous lattice heating observed here and with the electron diffraction experiments reported in Ref.~\onlinecite{Sciaini:2009qc} and is consistent with the diffusion of a cold plasma seen here.

As the plasma cools, it is difficult to reliably estimate the change in the plasma cooling rate: Taking an estimate of the electron density of states based on a Dirac model of the bands near the $L$-point would imply $\gamma_Q \sim T_p$ but an estimate of the density of states based on a parabolic band model (more appropriate for the holes near the $T$-point) would imply $\gamma_Q \sim T_p^{-1/2}$. In reality, we expect the temperature dependence of $\gamma_Q$ to be relatively weak, with the changes in the factors $\langle D_{avail}\rangle$ and $\hbar\omega/kT_p$ almost canceling each other as $T_p$ becomes smaller.

\subsection{Electron-hole recombination in the hot plasma}

For the hot plasma (with $kT_p \sim 0.5$~eV), assuming phonon-assisted electron-hole recombination between the electrons near the $L$-point and holes near the $T$-point in the Brillouin zone, we expect two factors to strongly suppress the recombination process, compared with the intra-band scattering responsible for momentum relaxation: (1) the small energy overlap of the conduction and valence bands implies that most electrons are at energies where they cannot scatter to the valence band (and similarly for holes scattering to the conduction band); (2) the density of states in the conduction-valence overlap region of the bands is much smaller than that typical of the hot carriers within the band, which determines the momentum relaxation time.
Including both factors (each approximately $10^{-3}$) suggests that the phonon-assisted electron-hole recombination is a factor of $\sim 10^{-6}$ smaller than the momentum relaxation time.
This rate is so small that this recombination mechanism is effectively closed off and it is likely that the actual electron-hole recombination rate is determined by other processes, most likely Auger recombination, which would be expected always to dominate at high plasma densities.

\subsection{Relaxation in the cool plasma}

Momentum relaxation, cooling and carrier recombination in the plasma that has cooled, close to room temperature, is determined by two different aspects of the density of states.
Only interband scattering between the electron and hole bands contributes to the electron-hole recombination rate, $\gamma_{e-h}$, whereas all bands contribute to the momentum relaxation rate.
Thus,  it is clear that $\gamma_{e-h} < \gamma_p$.
The momentum relaxation and the cooling rate of the lattice are still related by Eq.~\ref{heat_relax_rate}, but $T_p \approx T_l$ and the ratio $\gamma_Q/\gamma_p$ is of the order of 1/10.
It is more difficult to estimate the relative size of $\gamma_Q$ and the electron-hole recombination rate, $\gamma_{e-h}$, although one would expect them to be of similar order of magnitude. To determine which is larger would require a more precise knowledge of the electron-phonon matrix elements for intra-band and inter-band scattering than is currently known.

\section{Conclusions}\label{sec:conclusions}

We summarize the essential observations of the experiment:

1. Most of the excitation energy of the electron-hole plasma is transferred to the
lattice before the plasma becomes uniform in the films. We know this because
the residual $\Delta$R/R after 150 ps is much larger at the front face than at the
back face. This residual dR/R can only come from heating of the lattice.
This means that much more energy was transferred to the lattice at the front
of the film than at the back. (Heat conduction through the lattice is too
slow to significantly equilibrate the temperature in the film in 150 ps.)

2. The plasma becomes uniform in approximately 5 ps. This comes from an
analysis of the short time ($<$10 ps) back-side reflectivity signal, based on
a generic picture of a plasma diffusing from the front-side of the film, as shown in Fig. \ref{f:EnergyDensity} (a) using Eq.\ref{e:2decay}. This signal can only come from the electron-hole plasma since strain and heat
conduction from the front of the film are too slow to reach the back side of
the film in less than 50 ps.

3. The back-side reflectivity $\Delta$R/R decays with a decay time of approx 20 ps
during the 10-50 ps time. Low excitation measurements of $\Delta$R/R as a function of laser
fluence (front and back face) show that, in this density regime, $\Delta$R is
proportional to the plasma density within the optical absorption depth.
Hence the plasma density itself is decaying at this rate.

4. In the low-fluence regime, the size of the peak of the reflectivity
signal from the back of the film is consistent with a total number of electron-hole
pairs in the film, which has not decayed substantially from the number
originally created by optical absorption at the front face. Indeed, the
reflectivity signal is consistent with the same rate of decay of the total
number of electron-hole pairs as that observed during the 10-50 ps time range. (We
note that this is not true in the high-fluence regime.)

From 1 and 2, we know that the rate of heating of the lattice by the plasma
in the first 5 ps following excitation is substantially faster than 1/5 ps$^{-1}$.
This is consistent with the electron diffraction results of
\citet{Sciaini:2009qc}, which estimate that the lattice heating time, following
photoexcitation of very thin bismuth films,  is approximately 2 ps.

From 3 and 4 we find that, for the low-fluence regime, the decay rate of the
total number of electron-hole pairs in the film is not substantially
different over the first few picoseconds than that observed directly in the 10-50 ps
time range.

Our overall conclusion is that:
The decay of the plasma density is not determined primarily by cooling of
the plasma. Therefore, the chemical potential of the electrons and holes
(near the L and T points of the Brillouin zone, respectively) must be
different and this difference decays with a time-constant approximately
equal to 20 ps. The rate of equilibration of the chemical potentials is the
electron-hole recombination rate. We find this rate to be substantially
smaller than that indicated by simple extrapolation of the low-temperature
analysis of Lopez \cite{LopezPR1968} to room temperature.

In summary, we are able to separate experimentally the spatiotemporal dynamics of photoexcited carriers in bismuth from lattice heating and strain.  We measure ambipolar diffusivity in the range of $\sim$18-40 cm$^2$/s and carrier recombination times of $\sim$12--26 ps at low excitation.   We find evidence of rapid energy loss from the plasma to the lattice in both low and high-fluence excitation but without a corresponding decay in the plasma density after a few ps, thus demonstrating that the electron-hole recombination (occurring on the order 20 ps) is not determined by the plasma cooling on this time-scale. At the highest density our results cannot rule out the possibility of electron-hole recombination time of the order of a picosecond or less, with the plasma density being determined primarily by its temperature.

\begin{acknowledgments}
This work was supported in part by the U.S. DoE, Grants No.  DE-FG02-00ER1503, and from the NSF FOCUS physics frontier center. YJC and CU acknowledge the support from NSF-DMR-0604549. SF acknowledges the support by Science Foundation Ireland. We thank Roberto Merlin for use of the 250 kHz amplified laser and many useful discussions.
\end{acknowledgments}

\appendix
\section{Electron-phonon coupling and carrier relaxation}

With regard to electron-hole recombination, this analysis is similar to the approach discussed in Ref.\ \onlinecite{LopezPR1968}, Section V.A, but avoids the assumption of very low temperature. In addition, we discuss the rate of transfer of energy from the electronic plasma to the lattice vibration modes.  It is important to note that the band overlap between band 5 (valence) at $T$ and band 6 (conduction) at $L$ in Bi is approximately 38 meV.
Thus, the electron and hole gases at room temperature are in an intermediate regime between degenerate ($E_F >> kT$) and classical ($f(\epsilon) << 1$).
For the highly optically excited electron-hole plasma, with temperatures of the order of $kT = 0.5$~eV, the gas can be treated classically. In either case, the variation of the density of states is very substantial over the range of the $kT$ and the simplifications used by Lopez are not quantitatively valid.

We allow for the possibility that the lattice temperature $T_l$, which determines the level of phonon occupation, can be different from the plasma temperature $T_p$, which determines the electronic excitation in the bands.
We also allow for the chemical potential $\mu_h$ within the hole (valence) band to be different from the value $\mu_e$ for the electron (conduction) bands.  For simplicity, we assume that the electron and hole temperatures are equal.

More generally, we consider the scattering of electrons at temperature $T_p$ between two different bands, $1$ and $2$, with chemical potentials $\mu_1$ and $\mu_2$, by a phonon of frequency $\omega$ at temperature $T_l$. Thus, we assume that electronic states of energy $\epsilon$ in band $j$ have a probability of occupation,
\begin{equation}
f_j(\epsilon) = {1 \over e^{ ( \epsilon - \mu_j)/kT_p} + 1}, \qquad{\rm for~}j=1 {\rm ~or~}2.
\end{equation}
Let $g_1(\epsilon)$ and $g_2(\epsilon)$ be the electronic density of states in bands $1$ and $2$, respectively.
The phonon occupation number is assumed to be determined by the lattice temperature:
\begin{equation}
N(\hbar \omega) = { 1\over e^{\hbar\omega / kT_l} - 1}.
\end{equation}
We denote the electron-phonon coupling matrix element by
\begin{equation}
M = \langle \psi_{1,k+q} | {dH \over dQ_{q} } | \psi_{2,k} \rangle,
\end{equation}
where $H$ is the electronic single-particle Hamiltonian and $Q_{q}$ is the normalized phonon amplitude for wave vector $q$. We will assume that the $q$ and $k$ dependence of $M$ and $\omega_q$ is negligible for scattering between the band extrema $1$ and $2$.
Then, using the Fermi Golden Rule,  we obtain a rate of scattering from states in region $1$ to region $2$, due to phonon absorption:
\begin{eqnarray}
R_{a, 1\rightarrow 2} &=&  {2 \pi \over \hbar}|M|^2 N(\hbar\omega) \times  \\
&\int_{-\infty}^{+\infty}&
g_1(\epsilon) g_2(\epsilon+\hbar \omega)
f_1(\epsilon) [1-f_2(\epsilon+\hbar\omega) ] \; d\epsilon. \nonumber
\end{eqnarray}
The corresponding term involving phonon emission is:
\begin{eqnarray}
R_{e, 1\rightarrow 2} &=& {2 \pi \over \hbar}|M|^2 [N(\hbar\omega)+1] \times \\
&\int_{-\infty}^{+\infty}&
g_1(\epsilon) g_2(\epsilon-\hbar \omega)
f_1(\epsilon) [1-f_2(\epsilon-\hbar\omega) ] \; d\epsilon. \nonumber
\end{eqnarray}
Noting that
\begin{eqnarray}
&f_1(\epsilon)N(\hbar \omega) [1 - f_2(\epsilon+\hbar\omega)]
~=~ \exp\left[ { \mu_1 - \mu_2 \over kT_p} \right]\times  \\
&\exp\left[ { \hbar\omega \over kT_p} - { \hbar\omega \over kT_l} \right][1 -f_1(\epsilon) ] ( N(\hbar\omega) +1) f_2(\epsilon+\hbar\omega), \nonumber
\end{eqnarray}
we see that
\begin{equation}
R_{a, 1\rightarrow 2} =
\exp\left[ { \mu_1 - \mu_2 \over kT_p} \right] \exp \left[ { \hbar\omega \over kT_p} - { \hbar\omega \over kT_l} \right]
R_{e, 2\rightarrow 1}.
\end{equation}
Similarly,
\begin{equation}
R_{a, 2\rightarrow 1} =
\exp\left[ { \mu_2 - \mu_1 \over kT_p} \right] \exp \left[ { \hbar\omega \over kT_p} - { \hbar\omega \over kT_l} \right]
R_{e, 1\rightarrow 2}.
\end{equation}
Then the rate of loss of carriers from band $1$ (and corresponding gain in band $2$) is
\begin{eqnarray}
-{dn_1 \over dt}&=& {dn_2 \over dt} = R_{a, 1\rightarrow 2} - R_{e, 2\rightarrow 1}
                             - R_{a, 2\rightarrow 1} + R_{e, 1\rightarrow 2} \qquad  \nonumber \\
&=& \left\{1 -  \exp\left[ { \mu_2 - \mu_1 \over kT_p} + { \hbar\omega \over kT_l} - { \hbar\omega \over kT_p} \right]  \right\} R_{a, 1\rightarrow 2}  \nonumber \\
&+&
\left\{\exp\left[ { \mu_1 - \mu_2 \over kT_p} + { \hbar\omega \over kT_l} - { \hbar\omega \over kT_p} \right] -1  \right\} R_{a, 2\rightarrow 1}.
\end{eqnarray}
For many of the cases of interest to us in photo-excited room temperature bismuth, the term in the exponent is small and we get
\begin{eqnarray}
-{dn_1 \over dt} ~&\approx& \left[ { \mu_1 - \mu_2 \over kT_p} \right]
(R_{a, 1\rightarrow 2} + R_{a, 2\rightarrow 1})  \nonumber \\
~~~~~ &-& \left[ { \hbar\omega \over kT_l} - { \hbar\omega \over kT_p} \right] (R_{a, 1\rightarrow 2} - R_{a, 2\rightarrow 1}) .
\end{eqnarray}
The energy of one phonon is transferred from the plasma to the lattice for each phonon emission and vice versa for each phonon absorption. Thus, the rate of heat transfer from the plasma to the lattice associated with this scattering process is given by
\begin{eqnarray}
{dQ \over dt} &=& \hbar \omega \left[ R_{e, 1\rightarrow 2} + R_{e, 2\rightarrow 1} - R_{a, 1\rightarrow 2} - R_{a, 2\rightarrow 1} \right] \nonumber \\
&=& \hbar\omega \left[ R_{e, 1\rightarrow 2} - R_{a, 2\rightarrow 1}\right]   +
\hbar\omega \left[R_{e, 2\rightarrow 1} - R_{a, 1\rightarrow 2} \right] \nonumber \\
&=&
\hbar \omega \left\{\exp\left[ { \mu_2 - \mu_1 \over kT_p} + { \hbar\omega \over kT_l} - { \hbar\omega \over kT_p} \right]  - 1 \right\} R_{a, 1\rightarrow 2} \nonumber \\
 ~&+& ~
\hbar \omega \left\{\exp\left[ { \mu_1 - \mu_2 \over kT_p} + { \hbar\omega \over kT_l} - { \hbar\omega \over kT_p} \right] -1  \right\} R_{a, 2\rightarrow 1} \nonumber \\
&\approx&
\hbar \omega \left[ { \mu_2 - \mu_1 \over kT_p} + { \hbar\omega \over kT_l} - { \hbar\omega \over kT_p} \right]  R_{a, 1\rightarrow 2}  \nonumber \\
~&+& ~
\hbar \omega \left[ { \mu_1 - \mu_2 \over kT_p} + { \hbar\omega \over kT_l} - { \hbar\omega \over kT_p} \right]  R_{a, 2\rightarrow 1} \nonumber  \\
&=&
\hbar \omega \left[ { \mu_2 - \mu_1 \over kT_p} \right]  \left(R_{a, 1\rightarrow 2} - R_{a, 2\rightarrow 1} \right) \nonumber \\
~&+& ~
\hbar \omega \left[ { \hbar\omega \over kT_l} - { \hbar\omega \over kT_p} \right]  \left(R_{a, 1\rightarrow 2} + R_{a, 2\rightarrow 1} \right).
\label{e:dqdt}
\end{eqnarray}
The first term in the final expression of (\ref{e:dqdt})  is the heat transferred to the lattice due to electron-hole recombination and the second term is the heat due to ordinary heating without recombination. We see that the average energy transferred to the lattice per electron-hole pair recombined is less than the energy of one phonon:
\begin{equation}
\Delta E_{\rm recomb} = \hbar \omega  \left(  {R_{a, 1\rightarrow 2} - R_{a, 2\rightarrow 1} \over (R_{a, 1\rightarrow 2} + R_{a, 2\rightarrow 1} }  \right) < \hbar \omega.
\end{equation}
Examining the expressions for $R_{a, 1\rightarrow 2}$ and $R_{a, 2\rightarrow 1}$, it is clear that $R_{a, 1\rightarrow 2} \approx R_{a, 2\rightarrow 1}$ when the relative change in the densities of states $g_i$ and occupation factors $f_i$ is small for a change of energy of $\hbar \omega$. In this case, $\Delta E_{\rm recomb} << \hbar\omega$.

Treating the electrons and holes as classical, non-degenerate Fermi gases at temperature $T_p$, with densities $n_e$ and $n_h$, respectively, we have that
\begin{equation}
{\mu_v - \mu_c \over kT_p} = 2 {\Delta n_h \over n_h}
=  2 {\Delta n_e \over n_e}= 2{ \Delta n \over n}.
\end{equation}
(As long as we are in the non-degenerate gas regime, minor corrections to this result can be obtained from the Joyce-Dixon relation between chemical potential and density.)
We then have that
\begin{eqnarray}
{d \Delta n \over dt} &=& - \; 2 \Delta n \left({ R_{a, v\rightarrow c} + R_{a, c\rightarrow v} \over n}\right)
 \\
&~&  - \left[ { \hbar\omega \over kT_l} - { \hbar\omega \over kT_p} \right] (R_{a, v\rightarrow c} - R_{a, c\rightarrow v}) \nonumber
\end{eqnarray}
and we see that the recombination rate (i.e. the decay rate of $\Delta n$ when $T_l = T_p$)  is simply $\gamma_{n} =  2 (R_{a, v\rightarrow c} + R_{a, c\rightarrow v})/n$.

The plasma-lattice heat transfer rate allows us to estimate the cooling rate of the plasma, considering it as a classical gas of electrons and holes with a heat capacity per unit volume equal to $3 nk$.
Taking the above expression for $dQ/dt$ and allowing for scattering between multiple bands, including intraband scattering, we can find the rate of cooling of the hot plasma when it transfers heat to the lattice:
\begin{eqnarray}
{dT_p \over dt}
&=&
{\hbar \omega\over 3 nk} \sum_{i, j}\left[ { \mu_j - \mu_i \over kT_p} \right]  R_{a, i\rightarrow j}
\nonumber \\
&~&+ ~
{\hbar \omega \over 3nk} \left[ { \hbar\omega \over kT_l} - { \hbar\omega \over kT_p} \right]
\sum_{i, j} R_{a, i\rightarrow j}  \nonumber \\
&=&
{\hbar \omega\over 3 k} \sum_{i, j}\left[ { \mu_j - \mu_i \over kT_p} \right]
{R_{a, i\rightarrow j} \over n} \nonumber \\
&~&- ~
(T_p-T_l) \left[{\hbar \omega \over 3kT_p} \right] \left[ { \hbar\omega \over kT_l}\right]
\sum_{i, j} {R_{a, i\rightarrow j}  \over n}.
\end{eqnarray}
Thus, the plasma cooling rate (or carrier energy relaxation rate) is
\begin{eqnarray}
\gamma_Q &=&
\left[{\hbar \omega \over 3kT_p} \right] \left[ { \hbar\omega \over kT_l}\right]
\sum_{i, j} {R_{a, i\rightarrow j} \over n}  \\
&=& \left[{\hbar \omega \over 3kT_p} \right] \left[ { \hbar\omega \over kT_l}\right]
\sum_{i, j}  \gamma_{i \rightarrow j}
= \left[{\hbar \omega \over 3kT_p} \right] \left[ { \hbar\omega \over kT_l}\right]  \gamma_p,
\nonumber
\end{eqnarray}
where $\gamma_p$ is the average carrier scattering rate, corresponding to a carrier momentum relaxation rate. The average phonon energy in bismuth is of the order of 10 meV so that,  for the initial, hot photo-excited plasma, where
$3kT_p = \hbar \omega_{\rm photon} = 1.5$~eV and $kT_l = 25$~meV,
we find that the energy relaxation or plasma  cooling rate $\gamma_Q \approx \gamma_p / 375$.
For the final, cold (room temperature) plasma, $T_p = T_l$ and
\begin{equation}
\gamma_Q  = \left[ { \hbar \langle \omega \rangle \over kT_l } \right]^2 \gamma_p \approx {\gamma_p \over 6}.
\end{equation}
It is clear that the momentum relaxation rate determines the decay of electronic excitations of symmetry lower than the full crystal symmetry. Thus, the observed relaxation rate --- of the order of several femtosecond ---  for the decay of the $E_g$ force in \citet{Li2011} is consistent with a plasma cooling time (and associated lattice heating time) of a few picoseconds.

We note that, whereas the contribution to the electron-hole recombination rate arises only in the energy region where the conduction and valence bands overlap (plus approximately 10 meV for the energy of the phonon participating in the recombination process), the contribution to plasma-lattice heat transfer when the plasma is hotter than the lattice arises from all bands.  In particular, the optical phonons throughout the Brillouin zone will be effective in this heat transfer. For this heat transfer, we must consider intraband scattering, where band $1$ is the same as band $2$, as well as interband scattering. Thus, carrier recombination is strongly suppressed by the bottleneck in the density of states in the overlap region between the conduction and valence bands.
In the case of the initially hot plasma (with typical $kT_p \approx 0.5$~eV), the electrons and holes are spread in energy over a range of energies of the order of $500$~meV, whereas the overlap between the conduction and valence bands is only approximately $40$~meV.
Moreover, the density of electronic states to scatter to is much smaller near the band edges than it is 500 meV into the band.


\begin{thebibliography}{29}
\bibitem[{A1g()}]{A1gNote}
\bibinfo{note}{In principle the rapidly expanding plasma can also generate
  coherent phonons during roughly the first half phonon periods.}
\end{thebibliography}
\end{document}